\documentclass[12pt]{article}

\usepackage{amsmath}
\usepackage{latexsym}

\usepackage{bbm}
\usepackage{epsfig}

\newcommand{\be}{\begin{equation}}
\newcommand{\ee}{\end{equation}}
\newcommand{\ba}{\begin{eqnarray}}
\newcommand{\ea}{\end{eqnarray}}
\newcommand{\no}{\nonumber\\}

\allowdisplaybreaks

\begin{document}

\title{\normalsize \hfill UWThPh-2008-4 \\[8mm]
\LARGE The oblique parameters in multi-Higgs-doublet models}

\author{
\\
W.~Grimus,$^{(1)}$\thanks{E-mail: walter.grimus@univie.ac.at}
\ L.~Lavoura,$^{(2)}$\thanks{E-mail: balio@cftp.ist.utl.pt}
\ O.M.~Ogreid,$^{(3)}$\thanks{E-mail: omo@hib.no}
\ and P.~Osland$^{(4)}$\thanks{E-mail: per.osland@ift.uib.no}
\\*[5mm]
$^{(1)} \!$ \small
Fakult\"at f\"ur Physik, Universit\"at Wien \\
\small
Boltzmanngasse 5, 1090 Wien, Austria
\\*[2mm]
$^{(2)} \!$ \small
Universidade T\'ecnica de Lisboa
and Centro de F\'\i sica Te\'orica de Part\'\i culas \\
\small
Instituto Superior T\'ecnico, 1049-001 Lisboa, Portugal
\\*[2mm]
$^{(3)} \!$ \small
Bergen University College, Bergen, Norway
\\*[2mm]
$^{(4)} \!$ \small
Department of Physics and Technology, University of Bergen \\
\small
Postboks 7803, N-5020 Bergen, Norway
\\*[7mm]
}

\date{29 February 2008}

\maketitle

\vspace*{3mm}

\begin{abstract}
We present general expressions for the oblique parameters $S$,
$T$,
$U$,
$V$,
$W$,
and $X$ in the $SU(2) \times U(1)$ electroweak model
with an arbitrary number of scalar $SU(2)$ doublets,
with hypercharge $\pm 1/2$,
and an arbitrary number of scalar $SU(2)$ singlets.
\end{abstract}

\newpage

\section{Introduction}

\paragraph{Definition of the oblique parameters}
The oblique parameters are a useful way to parametrize
the effects of new physics (NP) on electroweak observables
when the following criteria are satisfied:
\begin{enumerate}
\item The electroweak gauge group is the standard $SU(2) \times U(1)$.
\item The NP particles have suppressed couplings
to the light fermions with which experiments are performed;
they couple mainly to the Standard-Model (SM) gauge bosons $\gamma$,
$Z^0$,
and $W^\pm$.
\item The relevant electroweak measurements
are those made at the energy scales $q^2 \approx 0$,
$q^2 = m_Z^2$,
and $q^2 = m_W^2$.
\end{enumerate}
When this happens,
the NP effects may be parametrized by only six quantities,
which were defined by Maksymyk \textit{et al.}~\cite{maksymyk},
following the work by various other authors~\cite{others1,others2},
as\footnote{We follow the convention
for the sign of the photon field  in~\cite{book}.}
\ba
\frac{\alpha}{4 s_W^2 c_W^2}\, S &=&
\frac{A_{ZZ} \left( m_Z^2 \right) - A_{ZZ} \left( 0 \right)}{m_Z^2}
-
\left. \frac{\partial A_{\gamma \gamma} \left( q^2 \right)}
{\partial q^2} \right|_{q^2=0}
+ \frac{c_W^2 - s_W^2}{c_W s_W}
\left. \frac{\partial A_{\gamma Z} \left( q^2 \right)}
{\partial q^2} \right|_{q^2=0},
\label{S} \\[1mm]
\alpha T &=&
\frac{A_{WW} \left( 0 \right)}{m_W^2}
-
\frac{A_{ZZ} \left( 0 \right)}{m_Z^2},
\label{T} \\[1mm]
\frac{\alpha}{4 s_W^2}\, U &=&
\frac{A_{WW} \left( m_W^2 \right) - A_{WW} \left( 0 \right)}{m_W^2}
- c_W^2\,
\frac{A_{ZZ} \left( m_Z^2 \right) - A_{ZZ} \left( 0 \right)}{m_Z^2}
\no[1mm] & &
- s_W^2
\left. \frac{\partial A_{\gamma \gamma} \left( q^2 \right)}
{\partial q^2} \right|_{q^2=0}
+ 2 c_W s_W
\left. \frac{\partial A_{\gamma Z} \left( q^2 \right)}
{\partial q^2} \right|_{q^2=0},
\label{U} \\[1mm]
\alpha V &=&
\left. \frac{\partial A_{ZZ} \left( q^2 \right)}
{\partial q^2} \right|_{q^2=m_Z^2}
- \frac{A_{ZZ} \left( m_Z^2 \right) - A_{ZZ} \left( 0 \right)}{m_Z^2},
\label{V} \\[1mm]
\alpha W &=&
\left. \frac{\partial A_{WW} \left( q^2 \right)}
{\partial q^2} \right|_{q^2=m_W^2}
- \frac{A_{WW} \left( m_W^2 \right) - A_{WW} \left( 0 \right)}{m_W^2},
\label{W} \\[1mm]
\frac{\alpha}{s_W c_W}\, X &=&
\left. \frac{\partial A_{\gamma Z} \left( q^2 \right)}
{\partial q^2} \right|_{q^2=0}
- \frac{A_{\gamma Z} \left( m_Z^2 \right)}{m_Z^2}.
\label{X1}
\ea
Here,
$\alpha = e^2 / \left( 4 \pi \right) = g^2 s_W^2 / \left( 4 \pi \right)$
is the fine-structure constant,
$s_W = \sin{\theta_W}$ and $c_W = \cos{\theta_W}$
are the sine and cosine,
respectively,
of the weak mixing angle $\theta_W$,
and the $A_{V V^\prime} \left( q^2 \right)$ are the coefficients
of $g^{\mu \nu}$ in the vacuum-polarization tensors
\be
\Pi^{\mu \nu}_{V V^\prime} \left( q \right) =
g^{\mu \nu} A_{V V^\prime} \left( q^2 \right)
+ q^\mu q^\nu B_{V V^\prime} \left( q^2 \right),
\ee
where $V V^\prime$ may be either $\gamma \gamma$,
$\gamma Z$,
$Z Z$,
or $W W$,
and $q = \left( q^\alpha \right)$
is the four-momentum of the gauge boson.

Our definition of the oblique parameters follows~\cite{maksymyk}
and allows for the case in which the NP scale
is not much higher than the Fermi scale:
it is not assumed that the $A_{V V^\prime} \left( q^2 \right)$
are linear functions of $q^2$.
The original definitions~\cite{others2} made that assumption and,
consequently,
there were only the three oblique parameters $S$,
$T$,
and $U$.

It is convenient to absorb into the oblique parameters
the prefactors on the left-hand sides of equations~(\ref{S})--(\ref{X1}),
by defining
\be\label{bardef}
\bar S \equiv \frac{\alpha}{4 s_W^2 c_W^2}\, S, \quad
\bar T \equiv \alpha T, \quad
\bar U \equiv \frac{\alpha}{4 s_W^2}\, U, \quad
\bar V \equiv \alpha V, \quad
\bar W \equiv \alpha W, \quad
\bar X \equiv \frac{\alpha}{s_W c_W}\, X.
\ee

It should be stressed that,
in the definition of an oblique parameter $O$,
a subtraction of the SM contribution should always be understood,
\textit{i.e.}
\be
O = \left. O \right|_\mathrm{NP} - \left. O \right|_\mathrm{SM}.
\ee
Therefore,
the $A_{V V^\prime} \left( q^2 \right)$
that we utilize in this paper are in reality
\be
A_{V V^\prime} \left( q^2 \right) =
\left. A_{V V^\prime} \left( q^2 \right) \right|_\mathrm{NP}
-
\left. A_{V V^\prime} \left( q^2 \right) \right|_\mathrm{SM}.
\ee
Thus,
the contributions to the $A_{V V^\prime} \left( q^2 \right)$
from loops of gauge bosons---including their
longitudinal components \textit{viz.}~the ``would-be''
Goldstone bosons---cancel.

The subtraction of the SM contributions
must also be used in the comparison of NP
with the precision data~\cite{Yao:2006px}.
One should note that,
in such a comparison,
one cannot~\cite{delta-rho-th} simultaneously determine from the data
the SM Higgs-boson mass and the oblique parameters $S$ and $T$.

Because of gauge invariance,
\be
A_{\gamma \gamma} \left( 0 \right)
=
A_{\gamma Z} \left( 0 \right)
=
0.
\label{zeros}
\ee
Therefore,
$X$ in equation~(\ref{X1}) may be rewritten as
\be
X =
\left.
\frac{\partial A_{\gamma Z} \left( q^2 \right)}{\partial q^2}
\right|_{q^2=0}
- \frac{A_{\gamma Z} \left( m_Z^2 \right)
- A_{\gamma Z} \left( 0 \right)}{m_Z^2}.
\label{X} 
\ee

\paragraph{The parameter $\Delta r$}
As a practical example of the application of the oblique parameters,
we may consider $\Delta r$,
defined by the relation~\cite{sirlin} (see also~\cite{hollik,denner})
\be
G_\mu = \frac{\pi \alpha}{\sqrt{2} m_W^2 s_W^2 \left( 1 - \Delta r \right)},
\label{dr}
\ee
where $s_W^2 \equiv 1 - m_W^2 / m_Z^2$.
The parameter $\Delta r$ contains the loop corrections
to the tree-level relation among $m_W$,
$m_Z$,
$\alpha$,
and the muon decay constant $G_\mu$.
Let us define
\be
\Delta r^\prime =
\left. \Delta r \right|_\mathrm{NP}
-
\left. \Delta r \right|_\mathrm{SM}.
\label{sub}
\ee
It is possible---provided that
the NP fields have suppressed couplings
to the light fermions involved in the measurements of $\alpha$,
$G_\mu$,
$m_Z$,
and $m_W$---to express $\Delta r^\prime$
in terms of the oblique parameters $S$,
$T$,
and $U$.
Indeed,
in that case $\Delta r^\prime$ originates solely
in modifications to the gauge-boson propagators,
\textit{viz.}~\cite{hollik}
\be
\Delta r^\prime =
\left. \frac{\partial A_{\gamma \gamma} \left( q^2 \right)}
{\partial q^2} \right|_{q^2 = 0}
+ \frac{A_{WW} \left( 0 \right) - A_{WW} \left( m_W^2 \right)}{m_W^2}
- \frac{c_W^2}{s_W^2} \left[
\frac{A_{ZZ} \left( m_Z^2 \right)}{m_Z^2}
- \frac{A_{WW} \left( m_W^2 \right)}{m_W^2} \right],
\ee
where equation~(\ref{zeros}) has been taken into account.
One then easily finds that~\cite{maksymyk}
\be
\Delta r^\prime = \frac{\alpha}{s_W^2} \left(
- \frac{1}{2}\, S + c_W^2 T + 
\frac{c_W^2 - s_W^2}{4 s_W^2}\, U \right).
\ee
This relation is useful for a comparison of any particular NP model
with the experimental data
\textit{viz.} the measured mass of the $W^\pm$ gauge bosons.
Indeed,
if one considers the measured values of $\alpha$,
$G_\mu$,
and $m_Z^2$ to constitute an experimental input
to the $SU(2) \times U(1)$ gauge theory,
then the relations~(\ref{dr}) and~(\ref{sub})
lead to the prediction of the $W^\pm$ mass
\be
m_W^2 = \left. m_W^2 \right|_\mathrm{SM}
\left( 1 + \frac{s_W^2}{c_W^2 - s_W^2}\, \Delta r^\prime \right).
\ee
Historically,
the parameter $\Delta r$ has played an important role
in the study of the SM;
for instance,
it has allowed the prediction of the top-quark mass
before the actual observation of that particle~\cite{top}.

\paragraph{Aim of this paper}
The purpose of this paper is
to give formulae for the six oblique parameters
at one-loop level\footnote{Since we are concerned with
the oblique parameters at only the one-loop level,
we are allowed to use throughout the tree-level relation
$m_W = c_W m_Z$.}
in an extension of the SM characterized by an enlarged scalar sector.
The scalar sector of the SM consists of only one $SU(2)$ doublet,
with hypercharge $1/2$.
In our NP model,
which we call the multi-Higgs-doublet-and-singlet model (mHDSM),
there is an arbitrary number of such scalar $SU(2)$ doublets,
together with an arbitrary number of scalar $SU(2)$ singlets
with arbitrary hypercharges.
It turns out that it is possible to derive simple,
closed formulae for the oblique parameters in the mHDSM,
in terms of only five functions of the masses of the scalar fields,
and of the matrix elements of only two mixing matrices.

In the mHDSM,
scalars with electric charges $0$ or $\pm 1$
are decoupled from scalars with any other electric charges.
Therefore, 
in section~\ref{model} we outline the mHDSM
in which all scalar fields have electric charges $0$ or $\pm 1$,
focussing especially on a general treatment
of the mixing of the scalars.
Section~\ref{results} contains our formulae
for the oblique parameters originating in that sector of the mHDSM.
In section~\ref{css} we develop the formulae
to the case in which $SU(2)$ singlets with electric charges
different from $0$ and $\pm 1$ are present in the mHDSM.
Section~\ref{conclusions} summarizes the findings of this paper.
A set of three appendices
explains some intermediate steps of our computations;
appendix~A compiles various relations
satisfied by the mixing matrices of the scalars,
appendix~B presents the needed Feynman integrals,
and appendix~C contains the functions of the scalar masses
which occur in the oblique parameters.

\section{The model} \label{model}

We consider an $SU(2) \times U(1)$ electroweak gauge model 
including $n_d$ scalar $SU(2)$ doublets $\phi_k$ with hypercharge $1/2$, 
$n_c$ complex scalar $SU(2)$ singlets $\chi_j^+$ with hypercharge $1$,
and $n_n$ real scalar $SU(2)$ singlets $\chi_l^0$ with hypercharge $0$:
\be
\label{fields}
\phi_k = \left( \begin{array}{c}
\varphi_k^+ \\ \varphi_k^0 \end{array} \right)
\quad (k = 1, 2, \ldots, n_d), 
\quad
\chi_j^+ \quad (j = 1, 2, \ldots, n_c), 
\quad
\chi_l^0 \quad (l = 1, 2, \ldots, n_n).
\ee
The neutral fields have vacuum expectation values (VEVs) 
\be
\left\langle 0 \left| \varphi_k^0 \right| 0 \right\rangle =
\frac{v_k}{\sqrt{2}},
\quad
\left\langle 0 \left| \chi_l^0 \right| 0 \right\rangle = u_l,
\ee
the $v_k$ being in general complex; 
the $u_l$ are real since the $\chi_l^0$ are real fields.
As usual,
we expand the neutral fields around their VEVs:
\be
\varphi_k^0 = \frac{v_k+ {\varphi_k^0}^\prime}{\sqrt{2}},
\quad
\chi_l^0 = u_l + {\chi_l^0}^\prime.
\label{expand}
\ee
Our treatment of the scalars was previously used in~\cite{Tpaper};
it is a generalization of the treatment in~\cite{grimus,GL02}.
The charged fields in~(\ref{fields})
can be expressed in terms of the charged mass eigenfields $S_a^+$ by 
\be
\label{charged}
\begin{array}{rcl}
\varphi_k^+ &=&
{\displaystyle \sum_{a=1}^n {\cal U}_{ka} S_a^+,}
\\*[3mm]
\chi_j^+ &=&
{\displaystyle \sum_{a=1}^n {\cal T}_{ja} S_a^+,}
\end{array}
\ee
with $n = n_d + n_c$.
Similarly,
the neutral fields in~(\ref{expand}) 
are linear combinations of the real neutral mass eigenfields $S_b^0$:
\be
\label{neutral}
\begin{array}{rcl}
{\varphi_k^0}^\prime &=&
{\displaystyle \sum_{b=1}^m {\cal V}_{kb} S^0_b,}
\\*[3mm]
{\chi_l^0}^\prime &=&
{\displaystyle \sum_{b=1}^m {\cal R}_{lb} S^0_b,}
\end{array}
\ee
with $m = 2n_d + n_n$.
The dimensions of the matrices in equations~(\ref{charged})
and~(\ref{neutral}) are
\be
{\cal U}\!: \; n_d \times n, \quad
{\cal T}\!: \; n_c \times n, \quad
{\cal V}\!: \; n_d \times m, \quad
{\cal R}\!: \; n_n \times m.
\ee
The matrices
\be
\label{mixing-matrices}
\tilde {\cal U} \equiv \left( \begin{array}{c} {\cal U} \\ 
{\cal T} \end{array} \right),
\quad
\tilde {\cal V} \equiv \left( \begin{array}{c}
\mbox{Re}\, {\cal V} \\ \mbox{Im}\, {\cal V} \\ {\cal R} \end{array}
\right)
\ee
are the diagonalizing matrices
for the mass-squared matrices of the charged and neutral scalars,
respectively.
The matrix $\tilde {\cal U}$ is $n \times n$ unitary,
the matrix $\tilde {\cal V}$ is $m \times m$ orthogonal.

Since we are dealing with a spontaneously broken
$SU(2) \times U(1)$ gauge theory, 
there are three unphysical Goldstone bosons,
$G^\pm$ and $G^0$,
which are ``swallowed'' by the $W^\pm$ and $Z^0$,
respectively,
to become their longitudinal components.
For definiteness we assign to them the indices $a=1$ and $b=1$,
respectively:
\be\label{goldstone}
S_1^\pm = G^\pm, \quad S_1^0 = G^0.
\ee
The masses of $G^\pm$
and of $G^0$---in a general 't~Hooft ($R_\xi$) gauge---are arbitrary
and unphysical:
they cannot appear in the final formula for any observable quantity.
We have checked,
by computing the oblique parameters in an arbitrary 't~Hooft gauge,
that all terms containing those masses do indeed cancel.

In the SM,
${\cal T}$ and ${\cal R}$ do not exist and ${\cal U} = \left( 1 \right)$,
${\cal V} = \left( i, 1 \right)$.

\section{The results} \label{results}

The parameter $T$ in the mHDSM was computed
in our previous paper~\cite{Tpaper},
where an extensive presentation of its derivation has been given.
Therefore,
we shall give here only the final result for that oblique parameter.

For all other five parameters,
all that one needs to calculate are the functions
\be
\frac{A_{V V^\prime} \left( q^2 \right)
- A_{V V^\prime} \left( 0 \right)}{q^2}, 
\label{diff}
\ee
for $V V^\prime = Z Z$,
$W W$,
or $\gamma Z$,
and $q^2 = m_{V^\prime}^2$,
and
\be
\left. \frac{\partial A_{VV} \left( q^2 \right)}
{\partial q^2} \right|_{q^2 = m_V^2}
- \frac{A_{VV} \left( m_V^2 \right) - A_{VV} \left( 0 \right)}{m_V^2},
\label{derdiff}
\ee
for $V=Z$ and $V=W$,
and also the derivatives of $A_{\gamma\gamma} \left( q^2 \right)$
and of $A_{\gamma Z} \left( q^2 \right)$ at $q^2 = 0$.

\begin{figure}
\begin{center}
\begin{tabular}{ccc}
\epsfig{file=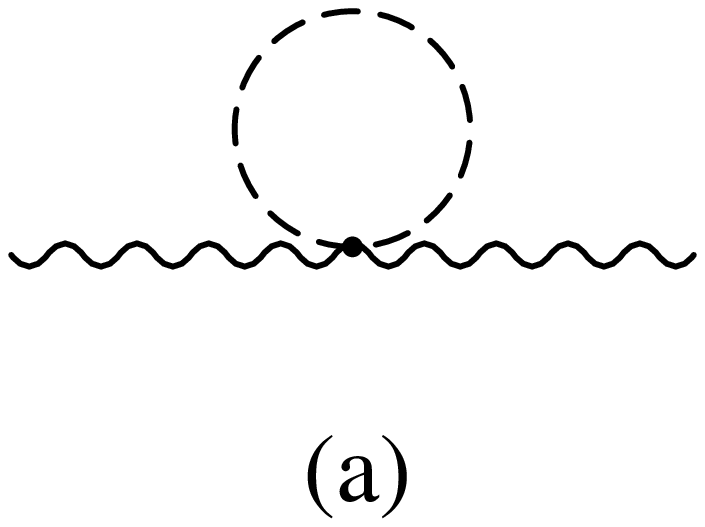,width=3.5cm} &
\epsfig{file=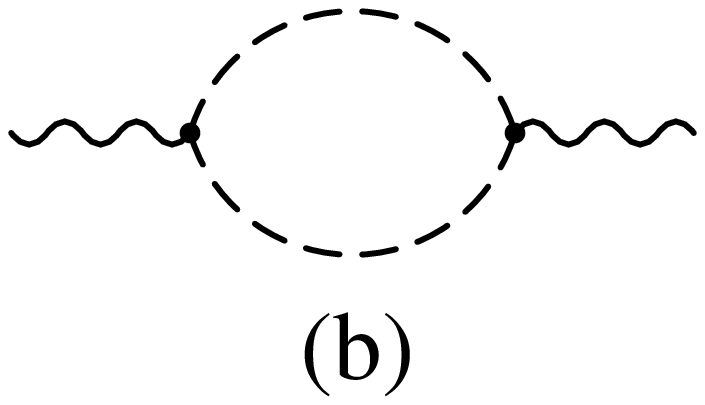,width=3.5cm} &
\epsfig{file=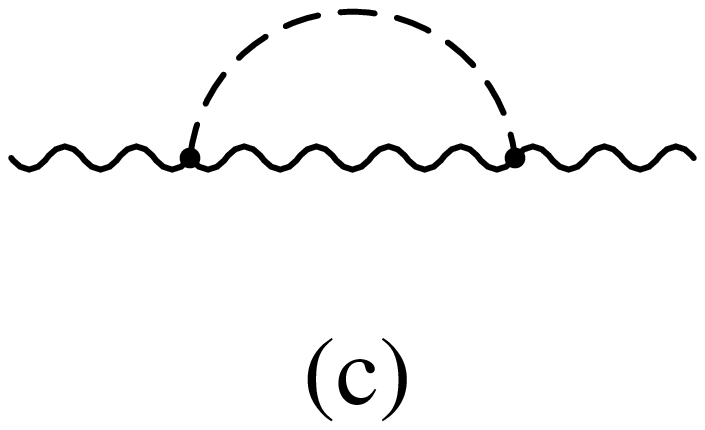,width=3.5cm} 
\end{tabular}
\end{center}
\caption{Three types of Feynman diagrams
occurring in the calculation of the vacuum polarizations.
\label{fig}}
\end{figure}

For a presentation of the Lagrangian of the mHDSM
in the physical basis of the scalars
we refer the reader to~\cite{Tpaper}.
We remark that the matrices ${\cal T}$ and ${\cal R}$
can be eliminated from the Lagrangian
by making use of the unitarity of $\tilde {\cal U}$
and the orthogonality of $\tilde {\cal V}$,
respectively,
in equation~(\ref{mixing-matrices}).
The Feynman diagrams which contribute to the vacuum polarizations
in the mHDSM are depicted in figure~\ref{fig}.\footnote{There are also
tadpole diagrams,
but they are irrelevant for the computation of the oblique parameters.} 
Diagrams of type~(a) are independent of $q^2$.
Therefore they affect only $T$.
It is also necessary to compute type (a) diagrams
if we want to demonstrate equation~(\ref{zeros}) in the mHDSM,
which we did;
except for this purpose,
type (a) Feynman diagrams are irrelevant
in the computation of $S$,
$U$,
$V$,
$W$,
and $X$,
and will henceforth not be considered in this paper.

There are no type~(c) diagrams for vacuum polarizations
involving either one or two photons---see
the mHDSM Lagrangian in~\cite{Tpaper}.
The type~(c) diagrams are only relevant for $A_{WW}$ and $A_{ZZ}$.

In the mHDSM all the oblique parameters,
except $T$,
are ultraviolet-finite after summation over all the Feynman diagrams
but {\em before} subtraction of the SM expression.
The parameter $T$,
on the other hand,
only becomes non-divergent
after the subtraction of the SM Higgs-boson loops,
as shown in~\cite{Tpaper}. 

\begin{figure}[htb]
\begin{center}
\epsfig{file=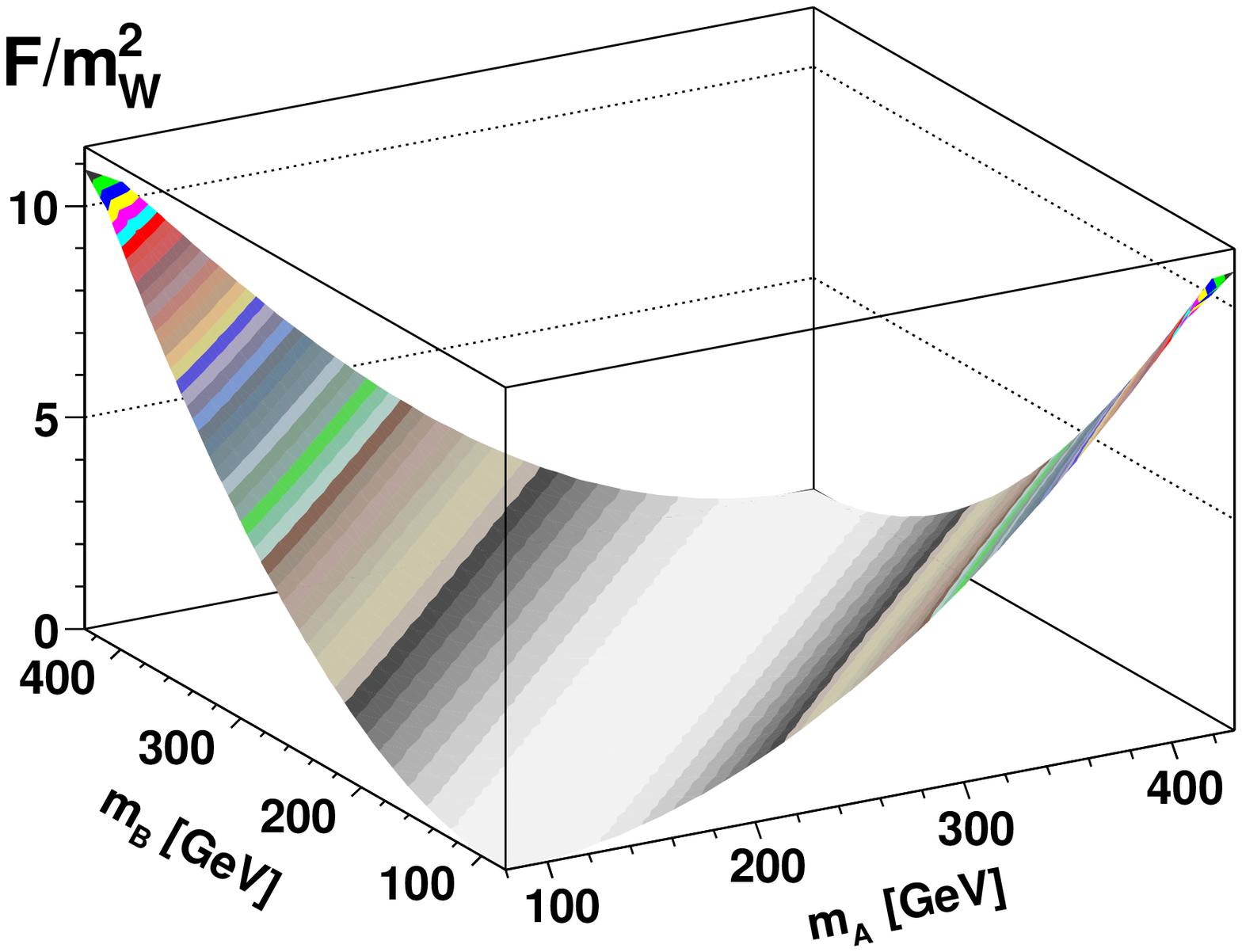,width=12cm}
\end{center}
\caption{$\left. F \left( m_A^2, m_B^2 \right) \right/ m_W^2$
vs.~$m_A$ and $m_B$.
\label{figF}}
\end{figure}

We begin by quoting the result for $T$ from~\cite{Tpaper}:
\ba
\bar T &=& \frac{g^2}{64 \pi^2 m_W^2} \left\{
\sum_{a=2}^n\, \sum_{b=2}^m\,
\left| \left( {\cal U}^\dagger {\cal V} \right)_{ab} \right|^2
F \left( m_a^2, \mu_b^2 \right)
\right. \nonumber
\\ & &
- \sum_{b=2}^{m-1}\, \sum_{b^\prime = b+1}^m\,
\left[ \mbox{Im} \left( {\cal V}^\dagger {\cal V} \right)_{b b^\prime} 
\right]^2
F \left( \mu_b^2, \mu_{b^\prime}^2 \right)
\nonumber
\\ & &
- 2\, \sum_{a=2}^{n-1}\, \sum_{a^\prime = a+1}^n\,
\left| \left( {\cal U}^\dagger {\cal U} \right)_{a a^\prime} \right|^2
F \left( m_a^2 , m_{a^\prime}^2 \right)
\nonumber
\\ & &
+ 3\, \sum_{b=2}^m\,
\left[ \mbox{Im} \left( {\cal V}^\dagger {\cal V} \right)_{1b} \right]^2
\left[
F \left( m_Z^2, \mu_b^2 \right) - F \left( m_W^2, \mu_b^2 \right)
\right]
\nonumber
\\ & & \left.
- 3 \left[
F \left( m_Z^2, m_h^2 \right) - F \left( m_W^2, m_h^2 \right)
\right]
\right\},
\label{finalT}
\ea
where $m_a$ denotes the mass of the charged scalars $S_a^\pm$
and $\mu_b$ denotes the mass of the neutral scalar $S_b^0$.
The second term in the right-hand side (RHS) of equation~(\ref{finalT})
contains a sum over all pairs of {\em different} physical neutral scalars,
\textit{i.e.}~$2 \le b < b^\prime \le m$;
similarly,
the third term in that RHS
contains a sum over all pairs of different {\em physical} charged scalars,
\textit{i.e.}~$2 \le a < a^\prime \le n$.
The last term in the RHS of equation~(\ref{finalT})
consists of the subtraction,
from the rest of $\bar T$,
of the SM result.
In that subtraction,
$m_h$ is the mass of the sole physical neutral scalar of the SM,
the Higgs particle.
The well-known~\cite{veltman-F} function $F$ is given by
\be
F \left( I, J \right) \equiv \left\{ \begin{array}{lcl}
\displaystyle{
\frac{I + J}{2} - \frac{I J}{I - J}\, \ln{\frac{I}{J}}
}
& \Leftarrow & I \neq J,
\\*[3mm]
0 & \Leftarrow & I = J.
\end{array} \right.
\label{funcF}
\ee
We depict this function in figure~\ref{figF}.

Next we write down the results for $\bar S$,
$\bar U$,
and $\bar X$:
\ba
\bar S &=&
\frac{g^2}{384 \pi^2 c_W^2} \left\{
\sum_{a=2}^n 
\left[ 2 s_W^2 - \left( {\cal U}^\dagger {\cal U} \right)_{aa} \right]^2
G \left( m_a^2, m_a^2, m_Z^2 \right)
\right. \no & &
+ 2 \sum_{a=2}^{n-1} \sum_{a^\prime = a+1}^n
\left| \left( {\cal U}^\dagger {\cal U} \right)_{a a^\prime} \right|^2
G \left( m_a^2, m_{a^\prime}^2, m_Z^2 \right)
\no & &
+ \sum_{b=2}^{m-1} \sum_{b^\prime = b+1}^m
\left[ \mbox{Im} \left( {\cal V}^\dagger {\cal V} \right)_{b b^\prime} 
\right]^2
G \left( \mu_b^2, \mu_{b^\prime}^2, m_Z^2 \right)
\no & &
- 2 \sum_{a=2}^n \left( {\cal U}^\dagger {\cal U} \right)_{aa} \ln{m_a^2}
+ \sum_{b=2}^m \left( {\cal V}^\dagger {\cal V} \right)_{bb} \ln{\mu_b^2}
- \ln{m_h^2}
\no & & \left.
+ \sum_{b=2}^m \left[ \mbox{Im} \left( {\cal V}^\dagger {\cal V} \right)_{1b} 
\right]^2
\hat G \left( \mu_b^2, m_Z^2 \right)
- \hat G \left( m_h^2, m_Z^2 \right)
\right\},
\label{finalS}
\\
\bar U &=&
\frac{g^2}{384 \pi^2} \left\{
\sum_{a=2}^n \sum_{b=2}^m
\left| \left( {\cal U}^\dagger {\cal V} \right)_{ab} \right|^2
G \left( m_a^2, \mu_b^2, m_W^2 \right)
\right. \no & &
- \sum_{a=2}^n
\left[ 2 s_W^2 - \left( {\cal U}^\dagger {\cal U} \right)_{aa} \right]^2
G \left( m_a^2, m_a^2, m_Z^2 \right)
\no & &
- 2 \sum_{a=2}^{n-1} \sum_{a^\prime = a+1}^n
\left| \left( {\cal U}^\dagger {\cal U} \right)_{a a^\prime} \right|^2
G \left( m_a^2, m_{a^\prime}^2, m_Z^2 \right)
\no & &
- \sum_{b=2}^{m-1} \sum_{b^\prime = b+1}^m
\left[ \mbox{Im} \left( {\cal V}^\dagger {\cal V} \right)_{b b^\prime} 
\right]^2
G \left( \mu_b^2, \mu_{b^\prime}^2, m_Z^2 \right)
\no & & 
+ \sum_{b=2}^m
\left[ \mbox{Im} \left( {\cal V}^\dagger {\cal V} \right)_{1b} \right]^2
\left[
\hat G \left( \mu_b^2, m_W^2 \right)
- \hat G \left( \mu_b^2, m_Z^2 \right)
\right] 
\no && \left.
\vphantom{\sum_{a=2}^n \sum_{b=2}^m
\left| \left( {\cal U}^\dagger {\cal V} \right)_{ab} \right|^2}
- \hat G \left( m_h^2, m_W^2 \right)
+ \hat G \left( m_h^2, m_Z^2 \right)
\right\},
\label{finalU}
\\
\bar X &=&
-\frac{g^2 s_W}{192 \pi^2 c_W} \sum_{a=2}^n
 \left[ 2 s_W^2 - \left( {\cal U}^\dagger {\cal U} \right)_{aa} \right]
G \left( m_a^2, m_a^2, m_Z^2 \right).
\label{finalX}
\ea
An explicit SM subtraction occurs in both $\bar S$ and $\bar U$.
In $\bar X$,
on the other hand,
such an explicit SM subtraction, 
showing the mass $m_h$ of the SM Higgs boson,
does not occur,
because $\bar X$ relates to $A_{\gamma Z} \left( q^2 \right)$,
and neutral particles---like the SM Higgs boson---do not couple
to the photon.
Still,
the SM subtraction has been performed in $\bar X$,
as elsewhere,
in order to remove,
from the sum over the charged scalars $S_a^\pm$,
the Goldstone-boson ($a=1$) term.
The explicit forms of the two functions $G \left( I, J, Q \right)$
and $\hat G \left( I, Q \right)$
are found in equations~(\ref{Gbar}) and~(\ref{Ghat})
of appendix~\ref{functions}.
Notice that,
contrary to what happens to the function $F \left( I, J \right)$
in equation~(\ref{funcF}),
the function $G \left( I, J, Q \right)$ does {\em not} vanish
when its first two arguments $I$ and $J$,
\textit{i.e.}~the squared masses of the two scalar particles
in the loop of a type~(b) diagram,
are equal.
In $\bar S$, the terms
\be
- 2 \sum_{a=2}^n \left( {\cal U}^\dagger {\cal U} \right)_{aa} \ln{m_a^2}
+ \sum_{b=2}^m \left( {\cal V}^\dagger {\cal V} \right)_{bb} \ln{\mu_b^2}
- \ln{m_h^2}
\label{termS}
\ee
are meaningful since
\be
- 2 \sum_{a=2}^n \left( {\cal U}^\dagger {\cal U} \right)_{aa} 
+ \sum_{b=2}^m \left( {\cal V}^\dagger {\cal V} \right)_{bb} - 1 = 0,
\ee
as can be verified by using equations~(\ref{sum1}),
(\ref{sum2}),
(\ref{UU11}),
and~(\ref{VV11});
therefore,
the terms~(\ref{termS}) are invariant
under a scaling of all the scalar masses by a common factor.

In order to write down formulae for $\bar V$ and $\bar W$
we need the functions $H \left( I, J, Q \right)$
and $\hat H \left( I, Q \right)$
in equations~(\ref{Jbar}) and~(\ref{Jhat}) of appendix~\ref{functions}.
We have
\ba
\bar V &=&
\frac{g^2}{384 \pi^2 c_W^2} \left\{
\sum_{a=2}^n
\left[ 2 s_W^2 - \left( {\cal U}^\dagger {\cal U} \right)_{aa} \right]^2
H\left( m_a^2, m_a^2, m_Z^2 \right)
\right. \no & &
+ 2 \sum_{a=2}^{n-1} \sum_{a^\prime = a+1}^n
\left| \left( {\cal U}^\dagger {\cal U} \right)_{a a^\prime} \right|^2
H \left( m_a^2, m_{a^\prime}^2, m_Z^2 \right)
\no & &
+ \sum_{b=2}^{m-1} \sum_{b^\prime = b+1}^m
\left[ \mbox{Im} \left( {\cal V}^\dagger {\cal V} \right)_{b b^\prime} 
\right]^2
H \left( \mu_b^2, \mu_{b^\prime}^2, m_Z^2 \right)
\no & & \left.
+ \sum_{b=2}^m \left[ \mbox{Im} \left( {\cal V}^\dagger {\cal V} \right)_{1b} 
\right]^2
\hat H \left( \mu_b^2, m_Z^2 \right)
- \hat H \left( m_h^2, m_Z^2 \right)
\right\},
\label{finalV} \\
\bar W &=&
\frac{g^2}{384 \pi^2} \left\{
\sum_{a=2}^n \sum_{b=2}^m
\left| \left( {\cal U}^\dagger {\cal V} \right)_{ab} \right|^2
H \left( m_a^2, \mu_b^2, m_W^2 \right)
\right. \no & & \left.
+ \sum_{b=2}^m \left[ \mbox{Im} \left( {\cal V}^\dagger {\cal V} \right)_{1b} 
\right]^2
\hat H \left( \mu_b^2, m_W^2 \right)
- \hat H \left( m_h^2, m_W^2 \right)
\right\}.
\label{finalW}
\ea

\section{Scalar singlets with electric charge other than $0, \pm 1$}
\label{css}

Scalar $SU(2)$ singlets with electric charge $Q \neq 0, \pm 1$
cannot mix with the scalars discussed in the previous two sections.
Moreover,
they couple to the photon and $Z^0$ but do not couple to the $W^\pm$.
Also,
there are no off-diagonal couplings
among different scalars with the same charge $Q$---for the Lagrangian,
see~\cite{Tpaper}.
Therefore, 
without loss of generality,
we may confine ourselves to a single scalar $SU(2)$ singlet,
with mass $m$ and electric charge $Q$.
In~\cite{Tpaper} we have shown that,
for such a scalar,
\be
\bar T = 0.
\ee
Since it does not couple to the $W^\pm$ gauge bosons,
$A_{WW} = 0$ and therefore
\be
\bar W = 0
\ee
too.

The fact that there are no type~(c) diagrams,
and that two identical scalars
occur in the loop of type~(b) diagrams,
greatly facilitates the computation of the other oblique parameters.
The results are
\ba
\bar S & = & \frac{Q^2 g^2 s_W^4}{96 \pi^2 c_W^2}\,
G \left( m^2, m^2, m_Z^2 \right), \\
\bar V & = & \frac{Q^2 g^2 s_W^4}{96 \pi^2 c_W^2}\,
H \left( m^2, m^2, m_Z^2 \right),
\ea
and the remaining oblique parameters are proportional to $\bar S$:
\ba
\bar U &=& - c_W^2 \bar S,
\\*[2mm]
\bar X &=& - \frac{c_W}{s_W}\, \bar S.
\ea

\section{Conclusions} \label{conclusions}

In this paper we have calculated the oblique
parameters---defined in a fashion appropriate
for new physics at a scale not necessarily much higher
than the Fermi scale~\cite{maksymyk}---in the standard
$SU (2) \times U(1)$ electroweak gauge theory
supplemented by an arbitrary number of scalar $SU(2)$ doublets
with hypercharge $1/2$
and scalar $SU(2)$ singlets with arbitrary hypercharges. 

We have found that the oblique parameters may be written
in terms of only two mixing matrices
${\cal U}$ and ${\cal V}$,
which parametrize the mixing of the charge-1 and charge-0 scalars,
respectively. 
These matrices take care simultaneously of the mixing of 
$SU(2)$-doublet scalars and $SU(2)$-singlet scalars, because 
${\cal U}$ is part of a larger unitary matrix $\tilde {\cal U}$,
while $\mbox{Re}\, {\cal V}$ and $\mbox{Im}\, {\cal V}$ are parts
of a larger real orthogonal matrix $\tilde {\cal V}$.
The expressions for the oblique parameters require only five functions 
$F \left( I, J \right) \left/ m_W^2 \right.$,
$G \left( I, J, Q \right)$,
$\hat G \left( I, Q \right)$,
$H \left( I, J, Q \right)$,
and $\hat H \left( I, Q \right)$,
of the squared scalar masses $I$ and $J$
and of the squared gauge-boson masses $Q = m_W^2$ or $Q = m_Z^2$.
We have depicted those functions in figures~2--5.

We note that for a model 
which has a compact spectrum,
in the sense that all the scalar masses---except
for the mass of the Higgs particle---are close together,
while they are all large compared to $m_Z$,
all those functions are usually smaller than 
$F \left( I, J \right) \left/ m_W^2 \right.$,
which grows like~\cite{barbieri}
$\left( \! \sqrt{I} - \sqrt{J} \right)^2$
and only appears in the parameter $T$.
This is one reason why 
in general we expect $T$ to be dominant
in the oblique corrections.\footnote{One exception
is the effective charge in atomic parity violation,
in which $T$ is multiplied by a very small factor
and $S$ may dominate~\cite{maksymyk}.
Another exception is the case
where only scalar singlets with electric charges
other than $0, \pm 1$ are present,
because then $T = 0$~\cite{Tpaper}.}
The other reason for the dominance of $T$
is the relatively large factor 
$\left. g^2 \right/ \left( 64 \pi^2 \alpha \right)
= \left. 1 \right/ \left( 16 \pi s_W^2 \right)$ contained in $T$---see
equations~(\ref{bardef}) and (\ref{finalT})---which multiplies 
$F \left( I, J \right) \left/ m_W^2 \right.$; 
the oblique parameters other than $T$ have smaller factors.

The functions $H \left( I, J, Q \right)$
and $\hat H \left( I, Q \right)$ tend to zero
when $I / Q$ and $J / Q$ grow.
The function $\hat G \left( I, Q \right)$
grows like $\ln{\left( I / Q \right)}$;
pure logarithms of the masses of the scalars
also appear in the expression for $S$,
and they may render that parameter relatively large
even if the masses of the new scalars are all equal.
The function $G \left( I, J, Q \right)$ is small for $I = J$
but it becomes sizable whenever $I$ and $J$ are quite far apart,
even if they are both much larger than $Q$,
\textit{i.e.}~than the electroweak scale.

\paragraph{Acknowledgements} W.G.\ thanks Stefan Dittmaier
for helpful discussions.
The work of L.L.\ was supported by the Portuguese
\textit{Funda\c c\~ao para a Ci\^encia e a Tecnologia}
through the project U777--Plurianual.
W.G.\ and L.L.\ acknowledge support from EU under the
MRTN-CT-2006-035505 network programme.

\appendix

\setcounter{equation}{0}
\renewcommand{\theequation}{A\arabic{equation}}

\section{Mixing-matrix relations}
\label{mixingmatrixrelations}

In this appendix we compile some useful relations
involving the mixing matrices ${\cal U}$ and ${\cal V}$.

The first set of relations follows
from the unitarity of the matrix $\tilde {\cal U}$
and the orthogonality of the matrix $\tilde {\cal V}$
defined in equations~(\ref{mixing-matrices}):
\ba
\sum_{b=1}^m
\left( {\cal U}^\dagger {\cal V} \right)_{ab}
\left( {\cal V}^\dagger {\cal U} \right)_{ba}
&=&
2 \left( {\cal U}^\dagger {\cal U} \right)_{aa},
\\
\sum_{a^\prime = 1}^n
\left( {\cal U}^\dagger {\cal U} \right)_{a a^\prime}
\left( {\cal U}^\dagger {\cal U} \right)_{a^\prime a}
&=&
\left( {\cal U}^\dagger {\cal U} \right)_{aa},
\\
\sum_{a=1}^n
\left( {\cal V}^\dagger {\cal U} \right)_{ba}
\left( {\cal U}^\dagger {\cal V} \right)_{ab}
&=&
\left( {\cal V}^\dagger {\cal V} \right)_{bb},
\\
\sum_{b^\prime = 1}^m
\left[ \mbox{Im} \left( {\cal V}^\dagger {\cal V} \right)_{b b^\prime} 
\right]^2
&=&
\left( {\cal V}^\dagger {\cal V} \right)_{bb}.
\label{sum10}
\ea
Furthermore,
since ${\cal U}$ is $n_d \times n$ and ${\cal V}$ is $n_d \times m$,
\ba
\label{sum1}
\sum_{a=1}^n \left( {\cal U}^\dagger {\cal U} \right)_{aa} &=& n_d, 
\\
\label{sum2}
\sum_{b=1}^m \left( {\cal V}^\dagger {\cal V} \right)_{bb} &=& 2 n_d.
\ea

A second set of relations follows from the convention
of placing the vectors pertaining to the Goldstone bosons
in the first columns of $\tilde {\cal U}$ and $\tilde {\cal V}$,
and also from the explicit form of those vectors,
namely~\cite{Tpaper,GL02}
\be
{\cal U}_{k1} = \frac{v_k}{v}, \quad {\cal V}_{k1} = i\, \frac{v_k}{v},
\quad \mbox{with} \
v \equiv \sqrt{ \sum_{k=1}^{n_d} \left| v_k \right|^2 } 
\simeq 246\ \mbox{GeV}.
\ee
Some ensuing relations are
\ba
\left( {\cal U}^\dagger {\cal U} \right)_{11} &=& 1,
\label{UU11}
\\
\left( {\cal V}^\dagger {\cal V} \right)_{11} &=& 1,
\label{VV11}
\\
\left( {\cal U}^\dagger {\cal V} \right)_{11} &=& i,
\\
\left( {\cal U}^\dagger {\cal U} \right)_{a1} &=& 0 \quad \Leftarrow a > 1,
\\
\left( {\cal U}^\dagger {\cal V} \right)_{a1} &=& 0 \quad \Leftarrow a > 1,
\\
\left( {\cal U}^\dagger {\cal V} \right)_{1b} &=& - \mbox{Im}
\left( {\cal V}^\dagger {\cal V} \right)_{1b} \quad \Leftarrow b > 1.
\label{rel2}
\ea

\setcounter{equation}{0}
\renewcommand{\theequation}{B\arabic{equation}}

\section{Feynman integrals}
\label{feynmanintegrals}

In this appendix we compute the Feynman integrals
which arise in type~(b) and type~(c) Feynman diagrams.

Diagrams of type~(b) lead to
\be
\label{IA}
i g^{\mu\nu} A \left( I, J, Q \right) =
\bar\mu^{4-d} \int \frac{\mathrm{d}^d k}{\left( 2 \pi \right)^d}\,
\int_0^1 \mathrm{d} x\,
\frac{4 k^\mu k^\nu}{\left( k^2 - \Delta + i\varepsilon \right)^2},
\ee
where $\bar\mu$ is the 't~Hooft mass,
$d$ the dimension of space--time,
and
\be
\Delta = Q x^2 + \left( J - I - Q \right) x + I.
\label{Delta}
\ee
In equation~(\ref{IA}),
$Q \equiv q^2$ is the squared four-momentum of the external gauge bosons,
$I$ and $J$ are the squared masses of the two scalar particles in the loop.
Then, 
\be
A \left( I, J, Q \right) =
\frac{1}{8 \pi^2} \int_0^1 \mathrm{d} x\,
\Delta \left( \mbox{div} - \ln{\Delta} \right) 
\ee
with
\be
\mbox{div} =
\frac{2}{4 - d} - \gamma + 1 + \ln{\left( 4 \pi \bar\mu^2 \right)},
\ee
$\gamma$ being the Euler--Mascheroni constant.
Explicitly,
\ba
A \left( I, J, Q \right) &=&
\frac{1}{8 \pi^2} \left\{
\vphantom{
\left[ \frac{\left( I - J \right)^2}{3 Q^2} - \frac{I + J}{Q}
\right] \frac{I - J}{4}\, \ln{\frac{I}{J}}
}
\left( \frac{I + J}{4} - \frac{Q}{12} \right)
\left( 2\, \mbox{div} - \ln{I} - \ln{J} \right)
\right. \no & &
+ \frac{2}{3} \left( I + J \right)
- \frac{5}{18}\, Q
- \frac{\left( I - J \right)^2}{6 Q}
\no & & \left.
+ \left[ \frac{\left( I - J \right)^2}{3 Q} - I - J
\right] \frac{I - J}{4 Q}\, \ln{\frac{I}{J}}
+ \frac{r}{12 Q^2}\, f \left( t, r \right) \right\}.
\label{AQ}
\ea
The function $f$ of
\be
\label{tr}
t \equiv I + J - Q
\quad \mbox{and} \quad
r \equiv Q^2 - 2 Q \left( I + J \right) + \left( I - J \right)^2
\ee
is given by
\be\label{f}
f \left( t, r \right) \equiv \left\{ \begin{array}{lcl}
{\displaystyle
\sqrt{r}\, \ln{\left| \frac{t - \sqrt{r}}{t + \sqrt{r}} \right|}
} & \Leftarrow & r > 0,
\\*[3mm]
0 & \Leftarrow & r = 0,
\\*[2mm]
{\displaystyle
2\, \sqrt{-r}\, \arctan{\frac{\sqrt{-r}}{t}}
} & \Leftarrow & r < 0.
\end{array} \right.
\ee
The absolute value in the argument of the logarithm takes effect only
if
\be
Q > \left( \sqrt{I} + \sqrt{J} \right)^2, 
\ee
in which case the vacuum polarization has an absorptive part;
in equation~(\ref{AQ}),
though,
only the dispersive part is given,
since it is the only one relevant for the oblique parameters.

In diagrams of type~(c) the internal particles
are a neutral scalar $S_b^0$ and a gauge boson $V$,
which may be either $V = Z^0$ if we are computing $A_{ZZ} \left( q^2 \right)$
or $V = W^\pm$ if we are computing $A_{WW} \left( q^2 \right)$.
We keep the notation of equation~(\ref{Delta})
but consider $I = \mu_b^2$ to be the squared mass of the neutral scalar
and $J = m_V^2$ to be the squared mass of the gauge boson.
Each vertex of a type~(c) diagram
contains one factor of the vector-boson mass $m_V$.
In a general 't~Hooft gauge,
the vector-boson propagator,
multiplied by $m_V^2$,
is
\be
\label{Vpropagator}
\frac{- m_V^2 g^{\mu \nu} + k^\mu k^\nu}{k^2 - m_V^2}
- \frac{k^\mu k^\nu}{k^2 - m_G^2},
\ee
where $m_G$ is the mass of the unphysical Goldstone boson,
\textit{i.e.} $m_G = m_1$ when $V = W^\pm$ and $m_G = \mu_1$ when $V = Z^0$.
The first term in~(\ref{Vpropagator})
leads to a Feynman integral different from the one of~(\ref{IA}):
\be
\label{IAbar}
i g^{\mu\nu} \bar A \left( I, J, Q \right) = 
\bar\mu^{4-d} \int \frac{\mathrm{d}^d k}{\left( 2 \pi \right)^d}\,
\int_0^1 \mathrm{d} x 
\frac{- 4 J g^{\mu \nu}}
{\left( k^2 - \Delta + i \varepsilon \right)^2}.
\ee
Consequently, 
\be
\bar A \left( I, J, Q \right) =
- \frac{J}{4 \pi^2} \int_0^1 \mathrm{d} x
\left( \mbox{div} - 1 - \ln \Delta \right).
\ee
Explicitly,
\ba
\bar A \left( I, J, Q \right) &=&
\frac{J}{8 \pi^2} \left[
- 2\, \mbox{div} + \ln{I} + \ln{J} - 2
+ \frac{I - J}{Q}\, \ln{\frac{I}{J}}
+ \frac{f \left( t, r \right)}{Q} \right].
\ea

Due to~(\ref{IA}) and~(\ref{Vpropagator}), 
the full expression for a type~(c) diagram is 
\be\label{full-c}
\bar A \left( I, J, Q \right)
+ A \left( I, J, Q \right)
- A \left( I, m_G^2, Q \right).
\ee
The third term in equation~(\ref{full-c})
contains the unphysical mass $m_G$ and,
therefore,
it cannot show up in the final results for the oblique parameters;
such terms either cancel internally during the computation
of the scalar contributions to the oblique parameters in the mHDSM,
or they cancel out when the SM contribution is subtracted
at the end of that computation.

In the limit $Q \to 0$ both $A \left( I, J, Q \right)$
and $\bar A \left( I, J, Q \right)$ are free from $Q^{-1}$ divergences
and can be written in terms of the function $F \left( I, J \right)$
in equation~(\ref{funcF})---see~\cite{Tpaper}.

\setcounter{equation}{0}
\renewcommand{\theequation}{C\arabic{equation}}

\section{Functions}
\label{functions}

In this appendix we derive the functions of the squared masses
which actually appear in the expressions for the oblique parameters.

\begin{figure}[htb]
\begin{center}
\epsfig{file=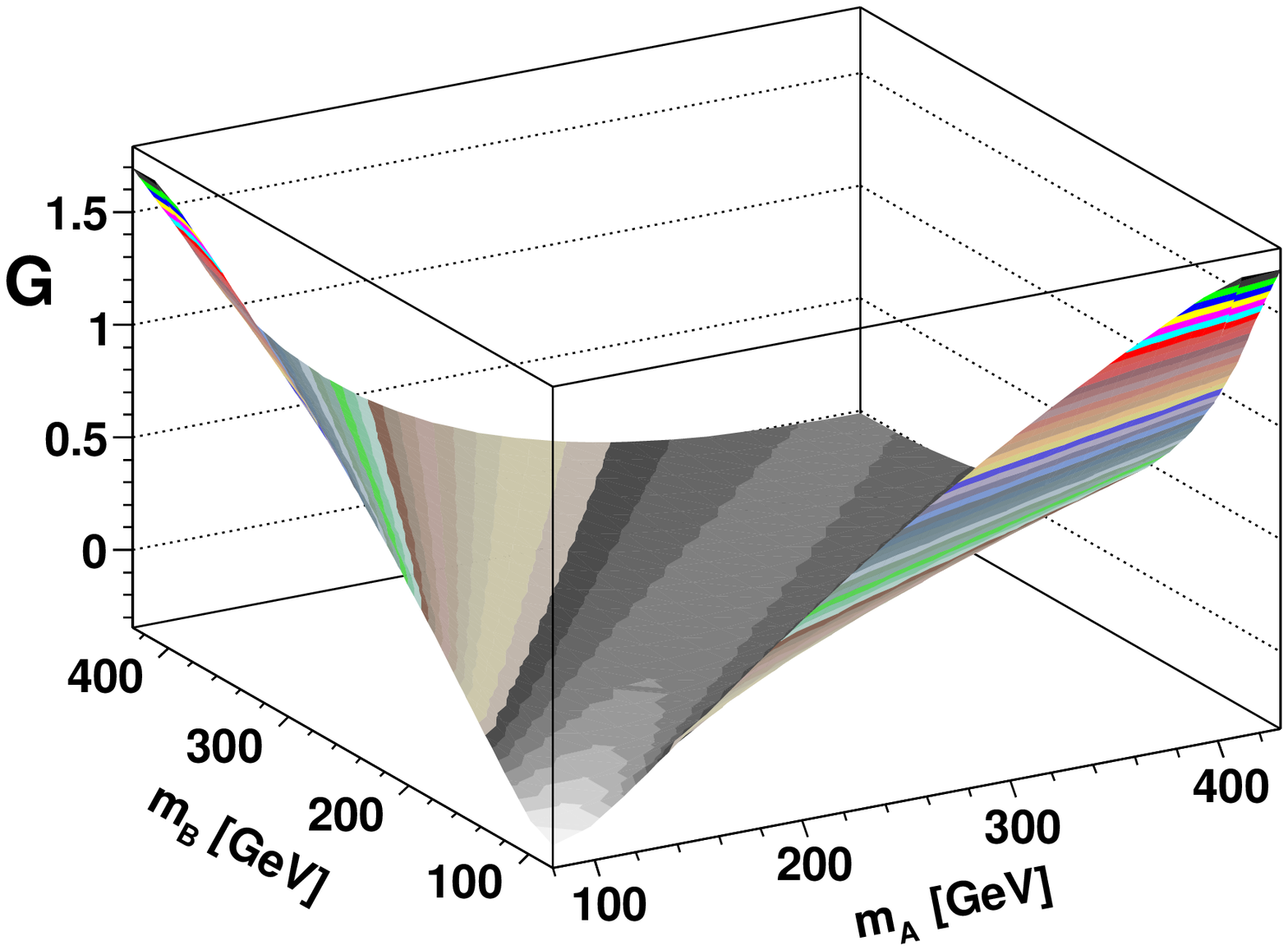,width=12cm}
\end{center}
\caption{$G(m_A^2,m_B^2,m_Z^2)$ vs.\ $m_A$ and $m_B$.
\label{fig-Gbar}}
\end{figure}

For terms of the form~(\ref{diff}) stemming from type~(b) diagrams,
we need
\be
\frac{A \left( I, J, Q \right) - A \left( I, J, 0 \right)}{Q} =
\frac{1}{96 \pi^2} \left[
2 - 2\, \mbox{div} + \ln{I} + \ln{J} + G \left( I, J, Q \right)
\right],
\ee
where
\ba
G \left( I, J, Q \right) &\equiv&
- \frac{16}{3}
+ \frac{5 \left( I + J \right)}{Q}
- \frac{2 \left( I - J \right)^2}{Q^2}
\no & &
+ \frac{3}{Q}
\left[ \frac{I^2 + J^2}{I - J}
- \frac{I^2 - J^2}{Q}
+ \frac{\left( I - J \right)^3}{3 Q^2} \right]
\ln{\frac{I}{J}}
+ \frac{r}{Q^3}\, f \left( t, r \right).
\label{Gbar}
\ea
This function is shown in figure~\ref{fig-Gbar} for $Q=m_Z^2$
and for a range of values
of $m_A \equiv \sqrt{I}$ and $m_B \equiv \sqrt{J}$.

In the case of type~(c) diagrams, we have to consider 
\be\label{diff-c}
\frac{\bar A \left( I, J, Q \right) - 
\bar A \left( I, J, 0 \right)}{Q} =
\frac{1}{8 \pi^2}\, \frac{J}{Q}\,
\tilde G \left( I, J, Q \right),
\ee
where
\be
\tilde G \left( I, J, Q \right) \equiv
- 2 + \left( \frac{I - J}{Q} - \frac{I + J}{I - J} \right) \ln{\frac{I}{J}}
+ \frac{f \left( t, r \right)}{Q}.
\ee
This is ultraviolet-finite because $\bar A \left( I, J, Q \right) $
has no ultraviolet divergences proportional to $Q$.
According to equation~(\ref{full-c}),
the full function which appears in the computation of type~(c) diagrams is
\ba
\hat G \left( I, Q \right) &\equiv&
G \left( I, Q, Q \right) + 12\, \tilde G \left( I, Q, Q \right)
\no &=&
- \frac{79}{3} + 9\, \frac{I}{Q} - 2\, \frac{I^2}{Q^2}
+ \left( - 10 + 18\, \frac{I}{Q} - 6\, \frac{I^2}{Q^2} + \frac{I^3}{Q^3}
- 9\, \frac{I + Q}{I - Q} \right) \ln{\frac{I}{Q}}
\no & &
+ \left( 12 - 4\, \frac{I}{Q} + \frac{I^2}{Q^2} \right)
\frac{f \left( I, I^2 - 4 I Q \right)}{Q}.
\label{Ghat}
\ea

\begin{figure}[htb]
\begin{center}
\epsfig{file=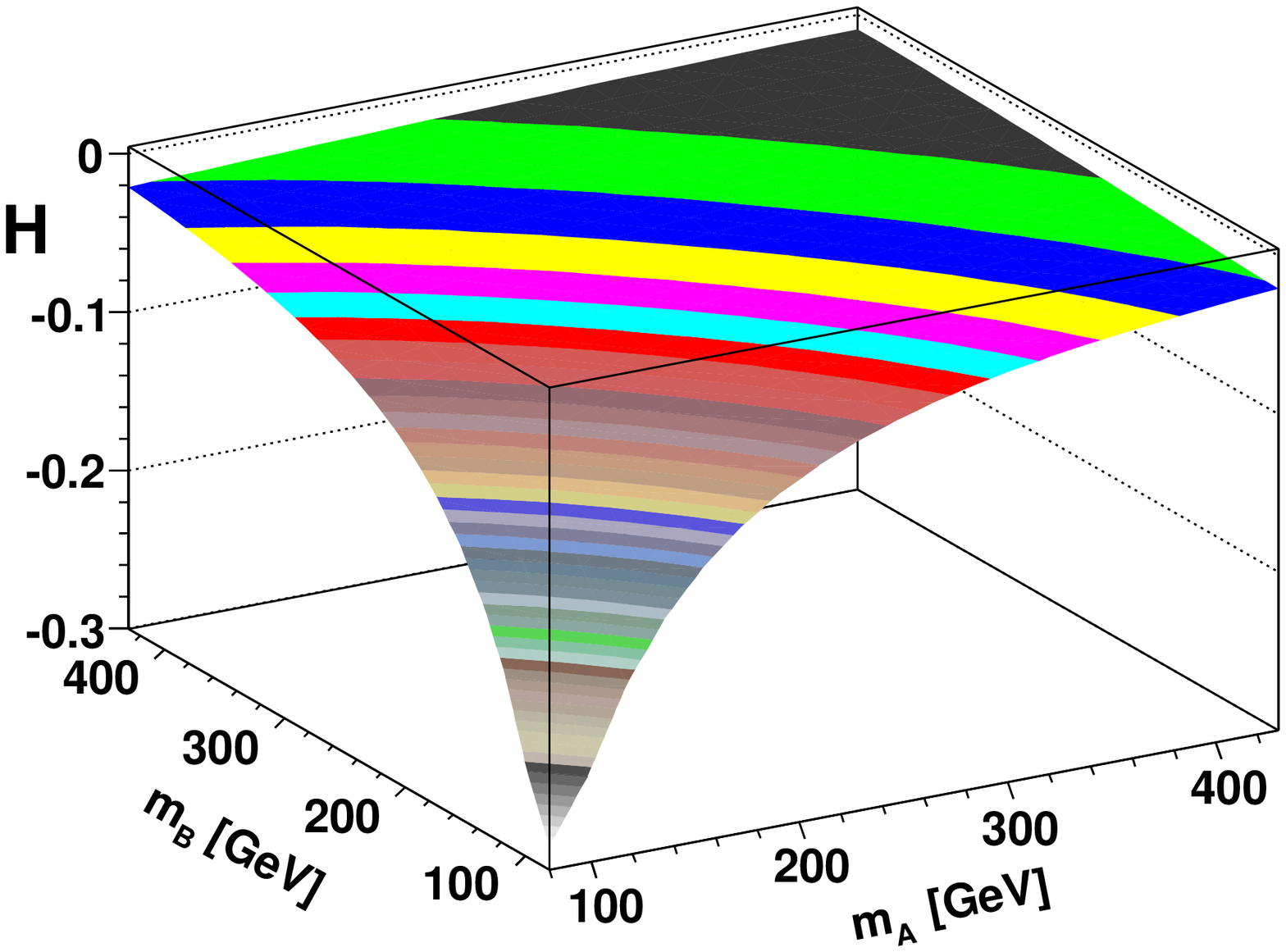,width=12cm}
\end{center}
\caption{$H(m_A^2,m_B^2,m_Z^2)$ vs.\ $m_A$ and $m_B$.
\label{fig-Hbar}}
\end{figure}

Proceeding to terms of the form~(\ref{derdiff}),
one has
\be
\frac{\partial A \left( I, J, Q \right)}{\partial Q}
- \frac{A \left( I, J, Q \right) - A \left( I, J, 0 \right)}{Q}
= \frac{1}{96 \pi^2}\, H \left( I, J, Q \right),
\label{jgurg}
\ee
which is ultraviolet-finite even for type~(b) diagrams.
Indeed,
\ba
H \left( I, J, Q \right) &\equiv&
2 - \frac{9 \left( I + J \right)}{Q}
+ \frac{6 \left( I - J \right)^2}{Q^2}
\no & &
+ \frac{3}{Q} \left[
- \frac{I^2 + J^2}{I - J}
+ 2\, \frac{I^2 - J^2}{Q}
- \frac{\left( I - J \right)^3}{Q^2} \right] \ln{\frac{I}{J}}
\no & &
+ \left[ I + J
- \frac{\left( I - J \right)^2}{Q} \right]
\frac{3\, f \left( t, r \right)}{Q^2}.
\label{Jbar}
\ea
This function is displayed in figure~\ref{fig-Hbar}.

Type~(c) diagrams give 
\be
\frac{\partial \bar A \left( I, J, Q \right)}{\partial Q}
- \frac{\bar A \left( I, J, Q \right) - \bar A \left( I, J, 0 \right)}{Q} =
\frac{1}{8 \pi^2}\, \frac{J}{Q}\,
\tilde H \left( I, J, Q \right),
\ee
where
\ba
\tilde H \left( I, J, Q \right) &\equiv&
4 + \left( \frac{I + J}{I - J} - 2\, \frac{I - J}{Q} \right) \ln{\frac{I}{J}}
\no & &
+ \frac{- Q^2 + 3 Q \left( I + J \right) - 2 \left( I - J \right)^2}
{r Q}\, f \left( t, r \right).
\ea
However,
in analogy to the function $\hat G \left( I, Q \right)$,
the function occurring in the full contribution of type~(c) diagrams is 
\ba
\hat H \left( I, Q \right)
& \equiv &
H \left( I, Q, Q \right) + 12\, \tilde H \left( I, Q, Q \right)
\no &=&
47 - 21\, \frac{I}{Q} + 6\, \frac{I^2}{Q^2}
+ 3 \left( 7 - 12\, \frac{I}{Q} + 5\, \frac{I^2}{Q^2} - \frac{I^3}{Q^3}
+ 3\, \frac{I + Q}{I - Q} \right) \ln{\frac{I}{Q}}
\no & &
+ 3 \left( 28 - 20\, \frac{I}{Q} + 7\, \frac{I^2}{Q^2} - \frac{I^3}{Q^3}
\right)
\frac{f \left( I, I^2 - 4 I Q \right)}{I - 4 Q}.
\label{Jhat}
\ea

\begin{figure}[htb]
\begin{center}
\epsfig{file=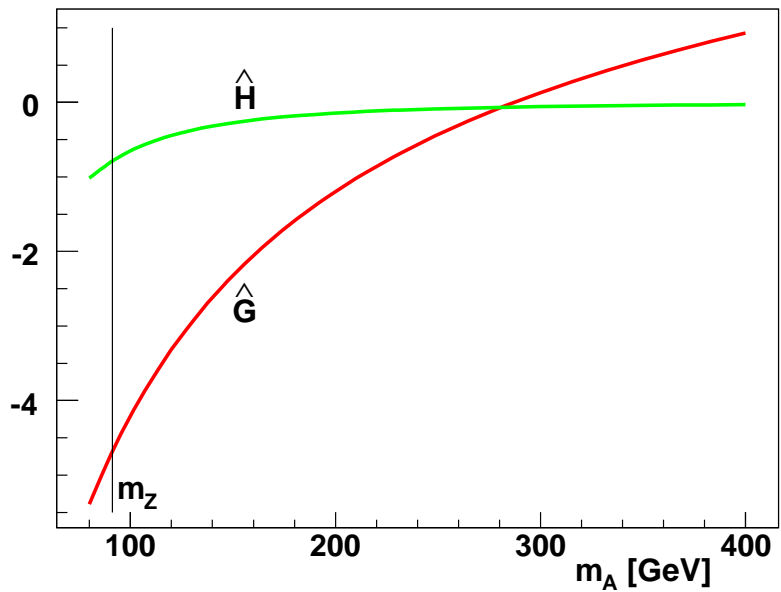,width=12cm}
\end{center}
\caption{$\hat G \left( m_A^2, m_Z^2 \right)$
and $\hat H \left( m_A^2, m_Z^2 \right)$ vs.~$m_A$.
\label{fig-GHhat}}
\end{figure}

The functions $\hat G \left( I, Q \right)$
and $\hat H \left( I, Q \right)$
are shown in figure~\ref{fig-GHhat},
for $Q = m_Z^2$ and a range of $m_A \equiv \sqrt{I}$.
Asymptotically,
with $\epsilon \equiv Q / I$,
\ba
\hat G \left( I, Q \right)
&=&
\left( - \frac{5}{3} + \ln{\frac{1}{\epsilon}} \right)
- \frac{17}{2}\,\epsilon
+ {\cal O} \left( \epsilon^2 \right),
\\
\hat H \left( I, Q \right)
&=&
-\frac{1}{2}\, \epsilon
-\frac{27}{10}\, \epsilon^2
+{\cal O} \left( \epsilon^3 \right).
\ea
Notice that,
when $\epsilon$ is very small,
{\it i.e.}~when the neutral-scalar masses
are much larger than the Fermi scale,
$\hat H \left( I, Q \right) \to 0$
but
$\hat G \left( I, Q \right)$ grows logarithmically like
$- \ln{\epsilon}$.

In $\bar S$,
$\bar U$,
and $\bar X$ there are contributions
from the derivatives with respect to $q^2$
of the photon self-energy,
and also of the mixed photon--$Z^0$ self-energy,
evaluated at $q^2 = 0$. 
Those self-energies only arise from type~(b) diagrams
and have two \emph{identical} charged scalars in the loop.
From equation~(\ref{AQ}),
one obtains the very simple expression
\be
\left. \frac{\partial A \left( I, I, Q \right)}{\partial Q} 
\right|_{Q = 0}
= \frac{1 - \mbox{div} + \ln{I}}{48 \pi^2},
\label{partialA0}
\ee
so that no new function beyond $F \left( I, J \right)$,
$G \left( I, J, Q \right)$,
$\hat G \left( I, Q \right)$,
$H \left( I, J, Q \right)$,
and $\hat H \left( I, Q \right)$ is needed for the oblique parameters.


\begin{thebibliography}{99}

\bibitem{maksymyk}
I.~Maksymyk, C.P.~Burgess, and D.~London,
\textit{Beyond S, T, and U},
\textit{Phys.\ Rev.}\ \textbf{D~50} (1994) 529
[hep-ph/9306267].

\bibitem{others1}
B.W.~Lynn, M.E.~Peskin, and R.G.~Stuart,
in \textit{Physics at LEP},
J.~Ellis and R.D.~Peccei eds. (CERN, Geneva, 1986);
\\
D.C.~Kennedy and B.W.~Lynn,
\textit{Electroweak radiative corrections with an effective Lagrangian:
Four-fermion processes},
\textit{Nucl. Phys.}\ \textbf{B~322} (1989) 1.

\bibitem{others2}
M.E.~Peskin and T.~Takeuchi,
\textit{A new constraint on a strongly interacting Higgs sector},
\textit{Phys.\ Rev.\ Lett.}\ \textbf{65} (1990) 964;
\\
G.~Altarelli and R.~Barbieri,
\textit{Vacuum-polarization effects of new physics
on electroweak processes},
\textit{Phys.\ Lett.}\ \textbf{B~253} (1991) 161;
\\
M.E.~Peskin and T.~Takeuchi,
\textit{Estimation of oblique electroweak corrections},
\textit{Phys.\ Rev.}\ \textbf{D~46} (1992) 381.
\\
G.~Altarelli, R.~Barbieri, and S.~Jadach,
\textit{Toward a model-independent analysis of electroweak data},
\textit{Nucl.\ Phys.}\ \textbf{B~369} (1992) 3
[erratum \textit{ibid.}\ \textbf{B~376} (1992) 444].

\bibitem{book}
G.C.~Branco, L.~Lavoura, and J.P.~Silva,
\textit{CP violation}
(Oxford University Press, New York, 1999),
chapter 11.

\bibitem{Yao:2006px}
W.M.~Yao \textit{et al.} (Particle Data Group),
\textit{Review of particle physics},
\textit{J.\ Phys.\ G (Particles and Fields)} \textbf{33} (2006) 1.

\bibitem{delta-rho-th}
J.~Erler and P.~Langacker in~\cite{Yao:2006px},
p.~119.

\bibitem{sirlin}
A.~Sirlin,
\textit{Radiative corrections in the $SU(2)_L \times U(1)$ theory:
A simple renormalization framework},
\textit{Phys.\ Rev.}\ \textbf{D~22} (1980) 971.

\bibitem{hollik}
W.~Hollik,
\textit{Radiative corrections in the Standard Model
and their role for precision tests of the electroweak theory},
\textit{Fortschr.\ Phys.}\ \textbf{38} (1990) 165.

\bibitem{denner}
M.~B\"ohm, A.~Denner, and H.~Joos,
\textit{Gauge theories of strong and electroweak interaction},
B.G.~Teubner, Stuttgart/Leipzig/Wiesbaden (2001),
chapter~4.6.

\bibitem{top}
For references see~\cite{Yao:2006px},
p.~526.

\bibitem{Tpaper}
W.~Grimus, L.~Lavoura, O.M.~Ogreid, and P.~Osland,
\textit{A precision constraint on multi-Higgs-doublet models},
arXiv:0711.4022 [hep-ph].

\bibitem{grimus}
W.~Grimus and H.~Neufeld,
\textit{Radiative neutrino masses in an $SU(2) \times U(1)$ model},
\textit{Nucl.\ Phys.}\ \textbf{B~325} (1989) 18.

\bibitem{GL02}
W.~Grimus and L.~Lavoura,
\textit{Soft lepton-flavor violation in a multi-Higgs-doublet seesaw model},
\textit{Phys.\ Rev.}\ \textbf{D~66} (2002) 014016 
[hep-ph/0204070].

\bibitem{veltman-F}
M.J.G.~Veltman,
\textit{Limit on mass differences in the Weinberg model},
\textit{Nucl.\ Phys.}\ \textbf{B~123} (1977) 89.

\bibitem{barbieri}
R.~Barbieri, L.J.~Hall, and V.S.~Rychkov,
\textit{Improved naturalness with a heavy Higgs:
An alternative road to LHC physics},
\textit{Phys.\ Rev.}\ \textbf{D~74} (2006) 015007
[hep-ph/0603188].


\end{thebibliography}
\end{document}